\documentclass[aps,prl,twocolumn,groupedaddress,showpacs]{revtex4}

\usepackage{graphicx}
\usepackage{dcolumn}
\usepackage{amsfonts}

\begin{document}


\title{Noise limits in the assembly of diffraction data}


\author{Veit Elser}
\affiliation{Department of Physics,
Cornell University,
Ithaca, NY 14853, USA}


\date{\today}

\begin{abstract}
We obtain an information theoretic criterion for the feasibility of assembling diffraction signals from noisy tomographs when the positions of the tomographs within the signal are unknown. For shot-noise limited data, the minimum number of detected photons per tomograph for successful assembly is much smaller than previously believed necessary, growing only logarithmically with the number of resolution elements of the diffracting object. We also demonstrate assembly up to the information theoretic limit with a constraint-based algorithm.
\end{abstract}


\maketitle

Arbitrarily high levels of noise can be tolerated when an unlimited number of measurements
are available and can be averaged to obtain the signal. A new challenge is introduced when
the signal is interrogated tomographically, that is, by means of multiple sections of the signal.
If the position of each tomograph within the signal is unknown, then each measurement must
have at least the minimum information required to position the tomograph, for signal averaging
to be feasible. We will refer to the signal processing demands posed by this scenario as
\textit{crypto-tomography}.

An instance of crypto-tomography is the assembly of diffraction data in the proposed x-ray free electron
laser (XFEL) investigations of single molecules \cite{Neutze}. The goal of these experiments is to obtain the 3D structure of
a molecule by algorithmically inverting to direct space the measurement of its continuous diffraction pattern.
Each XFEL measurement provides information about one 2D tomograph (an Ewald sphere) of the 3D
diffraction pattern. Since the molecular orientation in each measurement is random, and the total number
of photons collected small, the data assembly problem will test the noise limits of crypto-tomography. To help
assess the feasibility of these proposed experiments, we have investigated crypto-tomography
in the case of a weak, shot-noise limited signal. Because our approach is information theoretic \cite{Shannon}, the results
apply to any algorithm that aims to assemble and average noisy diffraction data. We conclude with some results obtained with an algorithm that is able to assemble data close to the information theoretic limit.

To better understand the theoretical issues, we introduce a minimal, three parameter model of crypto-tomography. A sample signal is shown in Figure 1 and consists of $N$ one-dimensional diffraction patterns
generated by $N$ one-dimensional objects. Each of the latter comprise $M$ independent, complex-valued resolution elements $\Psi_{m n}$. The diffraction signal is given by
\begin{equation}\label{signal}
w_n(\theta)=\left|\sum_{m=-M/2}^{M/2}\Psi_{m n}\,e^{i m\theta}\right|^2\quad n=1,\ldots,N\;,
\end{equation}
where $\theta$ is the single angle that specifies the position of the tomograph. For any $\theta$, the numbers
$w_n(\theta)$ are the time-integrated photon fluxes recorded at $N$ detector pixels. The third parameter of the model is the mean photon count per pixel, $\mu$. We will be interested in the limit of large $M$ and $N$. In this limit, the mean photon count is related to the statistics of the ensemble of resolution elements:
\begin{equation}\label{psinorm}
\langle |\Psi_{m n}|^2\rangle=\mu/M\;.
\end{equation}
A single exercise in crypto-tomography consists first in selecting one $M\times N$ set of resolution elements with the above statistics; this defines the correct diffraction signal. A noisy data set is then generated by repeatedly selecting a random $\theta$ and sampling $N$ photon counts $k_1$, $k_2$, etc. from Poisson distributions with means given by $w_1(\theta)$, $w_2(\theta)$, etc. Every data item thus consists of an $N$-tuple of photon counts. Given an unlimited number of such data, we are interested in determining the feasibility of reconstructing the original signal as a function of the model parameters $M$, $N$, and $\mu$.

Huldt \textit{et al.} \cite{Huldt} studied crypto-tomography from the perspective of \textit{classifying} the recorded photon counts. 
For the $N$-tuples of data in the model above, a decision is made if a pair $\{k_1(\theta_i),\ldots\}$ and $\{k_1(\theta_j),\ldots\}$ originated from different angles, $\theta_i\ne\theta_j$, or nearly the same angle, $\theta_i\approx\theta_j$ (on the scale $2\pi/M$ since $M$ is the highest frequency in the angular variation of the signal). This decision is based on the value of the cross-correlation
\begin{equation}
c(\theta_i,\theta_j)=\sum_{n=1}^{N}k_n(\theta_i) k_n(\theta_j)
\end{equation}
whose expected value distinguishes between counts derived from a common signal ($\theta_i\approx\theta_j$) or two independent signals \cite{Huldt}:
\begin{equation}
c_{i i}=\langle c(\theta_i,\theta_i)\rangle=2N\mu^2\quad c_{i j}=\langle c(\theta_i,\theta_j)\rangle=N\mu^2\;.
\end{equation}
The averages are with respect to Poisson distributions of photon counts with mean values given by diffraction signals $w$ with Wilson statistics, $p(w)=e^{-w/\mu}/\mu$.
A pair of $N$-tuples from different angles may be misidentified as originating from the same angle because the cross-correlation itself fluctuates. Evaluating the standard deviation $\sigma_{i j}$ of $c(\theta_i,\theta_j)$ one finds \cite{Huldt}
\begin{equation}
\sigma_{i j}^2 = N(\mu^2+4\mu^3+3\mu^4)\;.
\end{equation}
To avoid classification errors we must have $c_{i i}-c_{i j}\gg \sigma_{i j}$, which reduces to the statement
\begin{equation}
N\mu \gg \sqrt{N}
\end{equation}
in the limit of small $\mu$. This criterion, however, cannot be the fundamental limit since it makes no reference to $M$. As shown below, there is sufficient information for assembly with the much smaller number of detected photons given by (\ref{mincount}).
Finally, we demonstrate an assembly algorithm, not based on classification, that succeeds in this regime of high noise.

\begin{figure}
\scalebox{.65}{\includegraphics{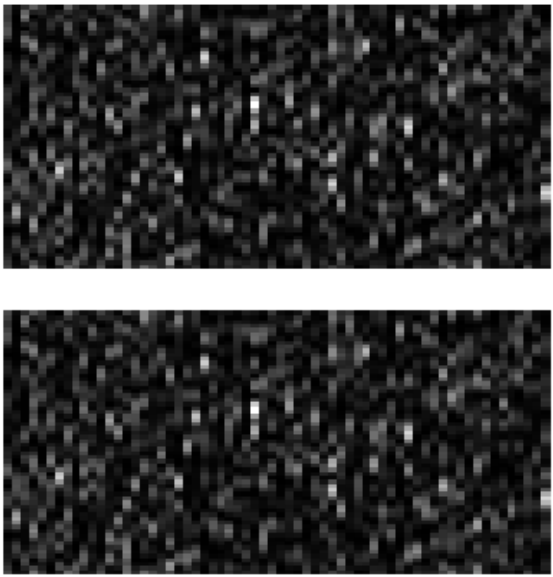}}
\caption{Top: A diffraction signal comprising 64 continuous one dimensional signals arranged side-by-side. The tomography angle $\theta$ varies vertically. Bottom: Reconstruction of the top signal from $10^4$ tomographic measurements (horizontal sections) of unknown $\theta$. Each measurement was a 64-tuple of photon counts with mean count $\mu=0.32$.\label{}}
\end{figure}

A criterion for the feasibility of crypto-tomography can be formulated in terms of
the information content of an abstract function $F$ that, given a
measured set of photon counts $K=\{k_1,\ldots,k_N\}$, checks consistency with the reconstructed signal $W(\theta)=\{w_1(\theta),\ldots,w_N(\theta)\}$. For the reconstruction to be unique, $F$ must be able to map fewer bits of input to a greater number of bits checked in the output. This is a restatement of the fact that $F$ has access to a unique reconstruction --- a positive source of information.

The number of bits in the output of $F$ is the mutual information \cite{Shannon} $I(K,W)$  associated with the joint probability distribution of photon counts and signals:
\begin{equation}
p(K,W)=p(W)\,\int\frac{d\theta}{2\pi}\, p\left(K|W(\theta)\right)\;.
\end{equation}
Here $p(W)$ is the prior distribution of signals, as specified by (\ref{signal}) and the statistics (\ref{psinorm}), and 
\begin{equation}\label{pktheta}
p\left(K|W(\theta)\right)=\prod_{n=1}^N \frac{w_n(\theta)^{k_n}}{k_n!}e^{-w_n(\theta)}
\end{equation}
is the Poisson distribution of photon counts at angle $\theta$ for signal $W$. The mutual information $I(K,W)$ gives the number of bits of information about $W$ obtained, on average, from each measurement $K$ at an unknown $\theta$.

Given a model reconstruction $W$, the number of independent consistency checks associated with a measurement $K$ has a size in bits given by the entropy of $\theta$ that remains, on average, after $K$ is known:
\begin{equation}
I(K,\theta)=H(\theta)-H(\theta|K)\;.
\end{equation}
This too is mutual information, now associated with the joint distribution $p(K,\theta)=p\left(K|W(\theta)\right)/2\pi$. Since we are interested in the uniqueness question for average case signals, $I(K,\theta)$ should be averaged over signals with the prior distribution $p(W)$.

The crypto-tomography criterion can now be written down explicitly:
\begin{equation}\label{criterion}
I(K,W)>\langle I(K,\theta)\rangle_W\;,
\end{equation}
where the mutual information expressions
\begin{widetext}
\[
I(K,W)=\int dW p(W) \sum_K p(K|W)\log{\frac{p(K|W)}{p(K)}}\qquad
\langle I(K,\theta)\rangle_W=\int dW p(W) \sum_K\int\frac{d\theta}{2\pi}\,p\left(K|W(\theta)\right)\log{\frac{p\left(K|W(\theta)\right)}{p(K|W)}}
\]
\end{widetext}
involve (\ref{pktheta}) and the marginal distributions $p(K|W)=\int p\left(K|W(\theta)\right)d\theta/2\pi$ and $p(K)=\int p(K|W)p(W) dW$. What follows is an analysis of what criterion (\ref{criterion}) implies about the parameters of our minimal model in the limit of large $M$ and $N$.

The definitions above imply that he sum
\begin{equation}\label{sum1}
I(K,W)+\langle I(K,\theta)\rangle_W=I\left(K,W(\theta)\right)
\end{equation}
corresponds to the mutual information associated with the photon counts $K$ and the signal $W(\theta)$ at a known angle $\theta$ of measurement. The joint distribution $p\left(K,W(\theta)\right)=p\left(K|W(\theta)\right)p\left(W(\theta)\right)$ involves the Wilson distribution of signal values:
\begin{equation}
dW p\left(W(\theta)\right)=\prod_{n=1}^N dw_n(\theta)\, e^{-w_n(\theta)/\mu}/\mu\;.
\end{equation}
Evaluating the mutual information we find $I\left(K,W(\theta)\right)=N I(\mu)$ where
\begin{equation}
I(\mu)=(1+\mu)\log{(1+\mu)}-\gamma\mu
-\sum_{k=2}^\infty\left(\frac{\mu}{1+\mu}\right)^k\log{k}\;.
\end{equation}
In the limit of interest, $\mu\to 0$, $I(\mu)\sim (1-\gamma)\mu$, with the result
\begin{equation}\label{sum2}
I\left(K,W(\theta)\right)\sim (1-\gamma)N\mu\;,
\end{equation}
or about $(1-\gamma)/\log{2}\approx 0.61$ bits per photon ($\gamma$ is Euler's constant).

We next evaluate $\langle I(K,\theta)\rangle_W$ in the limits of few and many detected photons $N\mu$. Since $N\mu$ is in either case large in the limit of large $N$, what matters is the relationship between $N\mu$ and the other parameter of the model, $M$. The few photon limit therefore corresponds to $N\mu$ fixed with $M\to\infty$. In this limit the complete prior distribution of signals is sampled by the process of sampling a particular, arbitrarily complex signal $W(\theta)$ at different $\theta$. The distribution $p(K|W)$ in the expression for $\langle I(K,\theta)\rangle_W$ thus can be replaced by the distribution $p(K)$ with the result
\begin{equation}\label{few}
\langle I(K,\theta)\rangle_W\sim I\left(K,W(\theta)\right)\qquad (M\to\infty)\;.
\end{equation}

The limit of many detected photons is an important point of reference, where the photon counts $K$ can be assigned a unique $\theta$ up to a width defined by a Gaussian distribution. Using the symmetry of the mutual information, we can write $I(K,\theta)=I(\theta,K)$ in the form
\begin{equation}
I(K,\theta)=\left< \int d\theta\,p(\theta|K)\log{2\pi p(\theta|K)}\right>_K\;.
\end{equation}
In the limit of large $N\mu$, the distribution of $\theta$ is a Gaussian centered at some $\theta_K$ with standard deviation $\sigma_K$:
\begin{equation}\label{gaussian}
\log{p(\theta|K)}\sim-\log{\sqrt{2\pi\sigma_K^2}}-\frac{(\theta-\theta_K)^2}{2\sigma_K^2}\;.
\end{equation}
The resulting mutual information is then given by
\begin{equation}\label{strong1}
I(K,\theta)\sim\left< \log{\sqrt{\frac{2\pi}{e\, \sigma_K^2}}}\right>_K\sim
\log{\sqrt{\frac{2\pi}{e}\left<\frac{1}{\sigma_K^2}\right>_K}}\;,
\end{equation} 
where the average over $K$ may be taken inside the logarithm because, as we shall see, $\sigma_K^{-2}$ is the sum of $N$ independent random terms.

Since $2\pi p(\theta|K)=p\left(K|W(\theta)\right)/p(K|W)$ depends on $\theta$ only through $p\left(K|W(\theta)\right)$, we have from (\ref{pktheta}) the equation
\begin{equation}
\log{p(\theta|K)}=\sum_{n=1}^N \left[ k_n \log{w_n(\theta)}-w_n(\theta)\right]+\mbox{const.}
\end{equation}
Using (\ref{gaussian}), we obtain
\begin{eqnarray}
\left<\frac{1}{\sigma_K^2}\right>_K&=&-\left<\left.\left(\frac{d}{d\theta}\right)^2\right|_{\theta=\theta_K} \log{p(\theta|K)}\right>_K\\
&=&\sum_{n=1}^N\left<\frac{w_n^\prime(\theta_K)^2}{w_n(\theta_K)}\right>_K\\
&=&\sum_{n=1}^N \left<\frac{w_n^\prime(\theta)^2}{w_n(\theta)}\right>_\theta\;,\label{strong2}
\end{eqnarray}
where the last step makes use of the fact that the distribution on $\theta_K$ associated with the distribution of $K$ is the uniform distribution.

\begin{figure}
\scalebox{.7}{\includegraphics{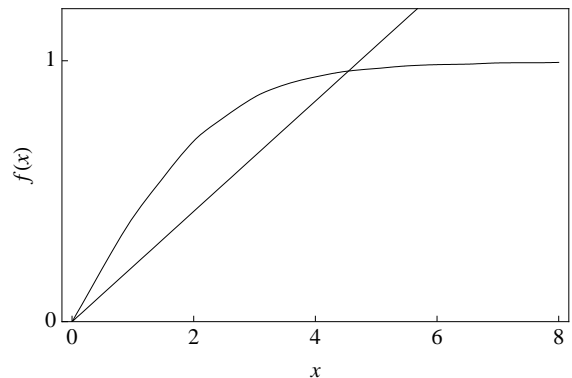}}
\caption{Plot of the mutual information scaling function $f(x)$ and its intersection with a line of slope $(1-\gamma)/2$.\label{}}
\end{figure}

The final step in the many photon limit of $\langle I(K,\theta)\rangle_W$ is to average (\ref{strong1}) over signals $W$ of the form (\ref{signal}). Since (\ref{strong2}) is again a sum of $N$ independent random terms, the average over $W$ may be taken inside the logarithm:
\begin{equation}\label{strong3}
\langle I(K,\theta)\rangle_W \sim
\log{\sqrt{\frac{2\pi}{e} N\left<\frac{w^\prime(0)^2}{w(0)}\right>_W}}\;.
\end{equation}
The remaining average can be expressed in terms of random variables $X$ and $Y$,
\begin{equation}
\frac{w^\prime(0)^2}{w(0)}=\frac{X}{X^\ast}(Y^\ast)^2+\frac{X^\ast}{X}(Y)^2 + 2|Y|^2
\end{equation}
where
\begin{eqnarray}
X&=&\sum_{m=-M/2}^{M/2} \Psi_m=\Psi_0+\sum_{m=1}^{M/2}(\Psi_m+\Psi_{-m})\\
Y&=&i\sum_{m=-M/2}^{M/2} m\Psi_m=i\sum_{m=1}^{M/2}m(\Psi_m-\Psi_{-m})\;.
\end{eqnarray}
Since $\Psi_m$ and $\Psi_{-m}$ are independent, so are $X$ and $Y$. Associated with the arbitrariness of the angle $\theta$ is the uniformity of the phases of the $\Psi_m$; consequently, the phases of $X$ and $Y$ are uniformly distributed and $\langle (Y^\ast)^2\rangle_W=\langle (Y)^2\rangle_W=0$. 
Using (\ref{psinorm}) for the third term we obtain
\begin{equation}\label{strong4}
\left<\frac{w^\prime(0)^2}{w(0)}\right>_W=2\langle |Y|^2\rangle_W\sim\frac{1}{6}M^2\mu
\end{equation}
in the limit of large $M$.

Combining (\ref{strong3}) and (\ref{strong4}), we obtain
\begin{equation}\label{many}
\langle I(K,\theta)\rangle_W \sim I_\infty=\log{\left(M\sqrt{\frac{2\pi}{6\, e}(N\mu)}\right)}\quad(N\mu\to\infty)
\end{equation}
for the many photon limit. We recognize $\log{M}$ as the scaling of the entropy of the tomography angle measured in speckle units of $2\pi/M$. Both of the limits (\ref{few}) and (\ref{many}) can be expressed in terms of a scaled photon count
\begin{equation}
x=N\mu/I_\infty\;,
\end{equation}
and a dimensionless scaling function
\begin{equation}\label{f(x)}
\langle I(K,\theta)\rangle_W \sim I_\infty\, f(x)
\end{equation}
behaving as $f(x)\sim (1-\gamma)x$ for small $x$ and $f(x)\sim 1$ for large $x$.
We have substantiated this claim by evaluating $\langle I(K,\theta)\rangle_W$ numerically for a large range of parameter values and find the results are consistent with a single function $f$ plotted in Figure 2.

\begin{figure}[!t]
\scalebox{.7}{\includegraphics{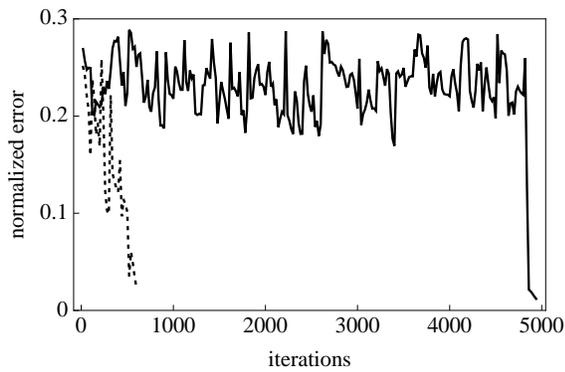}}
\caption{Normalized difference map error versus iteration in overconstrained ($\mu=0.32$) and underconstrained (\mbox{$\mu=0.25$}, dashed curve) crypto-tomography reconstructions.\label{}}
\end{figure}

Using (\ref{sum1}), (\ref{sum2}) and (\ref{f(x)}), the criterion (\ref{criterion}) takes the form
\begin{equation}
(1-\gamma)x>2 f(x)
\end{equation}
which, as shown in Figure 2, requires $x>4.5$. We therefore conclude that crypto-tomography for the three parameter model is feasible only when
\begin{equation}\label{mincount}
N\mu > 4.5 \log{\left(M\sqrt{\frac{2\pi}{6\, e}(N\mu)}\right)}\;.
\end{equation}

As an alternative to assembling a diffraction signal by first classifying its noisy tomographs, we propose using the tomographs as constraints on a \textit{de novo} reconstruction. A general constraint satisfaction algorithm may then be able to operate 
right at the information theoretic limits of feasibility. We now present results obtained with the iterative difference map algorithm \cite{diffmap} that support this claim. A description of the algorithm and more extensive results are given elsewhere.

To test the criterion (\ref{criterion}) we generated $10^4$ sets of photon counts from random tomographs taken from an instance of the three parameter model with \mbox{$M=16$} and $N=64$. Inequality (\ref{mincount}) then implies that crypto-tomography is possible only for mean photon counts \mbox{$\mu>0.26$}. 
An example of a successful reconstruction, for the case $\mu=0.32$, is shown in Figure 1. Figure 3 shows the corresponding difference map error \cite{diffmap}, a measure of the incompatibility of constraints at each iteration. An error that remains large during the search for the solution of a constraint satisfaction problem is an indication that the problem is overconstrained and that the solution will be unique. Correct reconstructions were obtained for $\mu$ as small as $0.29$, at which point the search became difficult and required very many iterations. For $\mu<0.26$ the behavior of the difference map error changed, decreasing to zero in few iterations (Fig. 3). This is consistent with the problem having become underconstrained, and in fact reconstructions obtained in this regime never agreed with the original diffraction signal. These observations suggest that the two forms of mutual information in the crypto-tomography criterion (\ref{criterion}) correspond, respectively, to the numbers of constraints and free variables in a constraint satisfaction problem.

\begin{acknowledgments}
The motivation for this work was prompted by discussions with Abbas Ourmazd. Support was provided by the Department of Energy grant DE-FG02-05ER46198.
\end{acknowledgments}

\bibliography{cryptotomo4}

\end{document}